\documentclass[aps,prl,reprint,superscriptaddress,floatfix,showpacs,longbibliography]{revtex4-1}
\usepackage{graphicx}
\usepackage{amssymb}
\usepackage{dcolumn}
\usepackage{bm}
\usepackage{epsfig}
\usepackage{verbatim}
\usepackage{amsmath}
\usepackage{amsfonts}
\usepackage{epic}
\usepackage{eepic}
\usepackage{color}
\usepackage{natbib}

\usepackage{url}
\definecolor{blue}{RGB}{34,30,127}
\usepackage[colorlinks=true,linkcolor=blue,urlcolor=blue,citecolor=blue]{hyperref}

\begin{document}

%\preprint{}

\title{The $N$ = 16 spherical shell closure in $^{24}$O}

% repeat the \author ..~\affiliation  etc.~as needed
% author.~Explanatory text should go in the []'s, actual e-mail

% \affiliation command.~The \affiliation command should follow the

\author{K.~Tshoo}
\email[]{tshoo99@snu.ac.kr}

%\thanks{}
\affiliation{Department of Physics and Astronomy, Seoul National University, Seoul 151-742, Korea.}
\author{Y.~Satou}
\author{H.~Bhang}
\author{S.~Choi}
\affiliation{Department of Physics and Astronomy, Seoul National University, Seoul 151-742, Korea.}
\author{T.~Nakamura}
\author{Y.~Kondo}
\author{S.~Deguchi}
\author{Y.~Kawada}
\author{N.~Kobayashi}
\author{Y.~Nakayama}
\author{K.N.~Tanaka}
\author{N.~Tanaka}
\affiliation{Department of Physics, Tokyo Institute of Technology, Tokyo 152-8551, Japan.}
\author{N.~Aoi}
\author{M.~Ishihara}
\author{T.~Motobayashi}
\author{H.~Otsu}
\author{H.~Sakurai}
\author{S.~Takeuchi}
\author{Y.~Togano}
\author{K.~Yoneda}
\author{Z.H.~Li} 
\affiliation{RIKEN Nishina Center, Saitama 351-0198, Japan.~}
\author{F.~Delaunay}
\author{J.~Gibelin}
\author{F.M.~Marqu\'es}
\author{N.A.~Orr}
\affiliation{LPC-Caen, ENSICAEN, Universit\'e de Caen, CNRS/IN2P3, 14050 Caen cedex, France.}
\author{T.~Honda}
\author{M.~Matsushita}
\affiliation{Department of Physics, Rikkyo University, Tokyo 171-8501, Japan.}
\author{T.~Kobayashi}
\affiliation{Department of Physics, Tohoku University, Aoba, Sendai, Miyagi 980-8578, Japan.}
\author{Y.~Miyashita}
\author{T.~Sumikama}
\author{K.~Yoshinaga} 
\affiliation{Department of Physics, Tokyo University of Science, Noda, Chiba 278-8510, Japan.}
\author{S.~Shimoura}
\affiliation{Center for Nuclear Study, University of Tokyo, Saitama 351-0198, Japan.}
\author{D.~Sohler}
\affiliation{Institute of Nuclear Research of the Hungarian Academy of Sciences, P.O.Box 51, H-4001 Debrecen, Hungary.~}
\author{T.~Zheng}
\author{Z.X.~Cao}
\affiliation{School of Physics and State Key Laboratory of Nuclear Physics and Technology, Peking University, Beijing 100871, China.}

%option in \documentclass).~\noaffiliation is required (may also be

\date{\today}

\begin{abstract}
The unbound excited states of the neutron drip-line isotope $^{24}$O have been investigated via the $^{24}$O($p$,$p'$)$^{23}$O+$n$ reaction in inverse kinematics at a beam energy of 62 MeV/nucleon.~The decay energy spectrum of $^{24}$O$^*$ was reconstructed from the momenta of $^{23}$O and the neutron.~The spin-parity of the first excited state, observed at $E_{\rm x}$ = 4.65 $\pm$ 0.14 MeV, was determined to be  $J^{\pi}$ = $2^+$ from the angular distribution of the cross section.~Higher-lying states were also observed.~The quadrupole transition parameter $\beta_{2}$ of the 2$_1^+$ state was deduced, for the first time, to be 0.15 $\pm$ 0.04.~The relatively high excitation energy and small $\beta_{2}$ value are indicative of the $N$ = 16 shell closure in $^{24}$O.
\end{abstract}


%\keywords{}

\maketitle
%\section{}
%\subsection{}
%\subsubsection{}
% If in two-column mode, this environment will change to single-column
% format so that long equations can be displayed.~Use
%\begin{widetext}
%\end{widetext}

%Magic number is a unique feature of a finite Fermionic quantum system.~For atomic nuclei, recent findings of new magic numbers and disappearance of the traditional well-known magic numbers in neutron-rich nuclei, called shell evolution, have attracted intense interests since it has the fundamental implication not only for the magicity itself in nuclei but also for understanding the cosmic abundance distribution.~For instance, the low excitation energies ($E_{\rm x}$) and large quadrupole transition parameters ($\beta_{2}$) of the first 2$^+$ states in $^{12}_4$Be$_{8}$~\cite{Iwasaki481,Navin00} and $^{32}_{12}$Mg$_{20}$~\cite{Motobayashi95} point to the disappearance of the magic numbers $N$ = 8 and 20.~Higher in mass, the recent observation of a low-lying 2$_1^+$ state in $^{42}_{14}$Si$_{28}$ provided evidence for the ``melting" of the $N$ = 28 shell closure~\cite{Bastin07}.

Magic numbers are a unique feature of finite Fermionic quantum systems.~In the case of atomic nuclei, experiments over the last decade or so have shown that the well-known magic numbers (2, 8, 20, 28, $\textellipsis$) seen to occur in stable nuclei often disappear and are replaced by new ones as the neutron or proton drip lines are approached~\cite{Sorlin08}.~For instance, the low excitation energies ($E_{\rm x}$) and large quadrupole transition parameters ($\beta_{2}$) of the first 2$^+$ states in $^{12}_4$Be$_{8}$~\cite{Iwasaki481} and $^{32}_{12}$Mg$_{20}$~\cite{Motobayashi95} point to the disappearance of the magic numbers $N$ = 8 and 20.~Higher in mass, the recent observation of a low-lying 2$_1^+$ state in $^{42}_{14}$Si$_{28}$ provided evidence for the ``melting" of the $N$ = 28 shell closure~\cite{Bastin07}.

In the oxygen isotopes, new shell closures at $N$ = 14~\cite{Thirolf00, Stanoiu04, Becheva06} and 16~\cite{Ozawa00, Rituparna01, Otsuka01, Otsuka05,BrownRichter, Otsuka10,Hoffman08, Hoffman09, Kanungo09} have been proposed.~A high excitation energy and low $B(E2)$ value of the $2_1^+$ state of $^{22}$O was determined via Coulomb excitation~\cite{Thirolf00}, which is sensitive to the charge distribution.~Proton inelastic scattering on $^{22}$O, which is sensitive to both the proton and neutron distributions, was studied by Becheva \textit{et al.}~\cite{Becheva06} and a small quadrupole transition parameter $\beta_{2}$(0$^+_{\rm g.s.}$$\rightarrow$$2_1^+$) = 0.26 $\pm$ 0.04 was reported.~Taken together these results indicate the near spherical character of $^{22}$O and a sizable gap at $N$ = 14.~Recently, the low-lying level structure of $^{24}$O was studied via proton knock-out from $^{26}$F by Hoffman \textit{et al.}~\cite{Hoffman09}.~They reported an even higher excitation energy ($E_{\rm x}$ = 4.72 $\pm$ 0.11 MeV) for the first excited state of $^{24}$O than that of $^{22}$O (3.20 $\pm$ 0.01 MeV)~\cite{Stanoiu04} suggesting, as proposed by Brown and Richter~\cite{BrownRichter}, a shell closure at $N$ = 16.~It should be noted, however, that the spin-parity assignments for the $^{24}$O excited states were only based on a comparison with the predicted shell-model energies. Otsuka \textit{et al.}~have investigated theoretically the structural evolution of the oxygen isotopes with increasing neutron number ($N$) and attributed the development of the shell closure at $N$ = 16 to the strong neutron-proton tensor interaction~\cite{Otsuka01,Otsuka05}.

%In this Letter we report on the first spectroscopic study of $^{24}$O by proton inelastic scattering.~In addition to the excitation energies of the states populated, the well-known character of proton inelastic scattering enables us to determine the spin-parity and the quadrupole transition parameter ($\beta_{2}$) of the first 2$^+$ state.~As described below, we determined the spin-parity and the $\beta_{2}$ value for the state at $E_{\rm x}$ =  4.65 $\pm$ 0.14 MeV to be 2$^+$ and 0.15 $\pm$ 0.04, respectively.~The small value of $\beta_{2}$ is indicative of the spherical closed-shell character of $^{24}$O.~A comparison of the $E_{\rm x}(2_1^+)$ and $\beta_{2}$ for the chain of oxygen isotopes shows strong evidence for a large shell gap at $N$ = 16.

In this Letter we report on the first spectroscopic study of $^{24}$O by proton inelastic scattering.~In addition to the excitation energies of the states populated, the well-known character of proton inelastic scattering also permits the spins-parities, as well as the quadrupole transition parameter ($\beta_{2}$) of the first 2$^+$ state to be deduced.~As described below, we have been able to provide a firm 2$^+$ assignment for the state at $E_{\rm x}$ =  4.65 $\pm$ 0.14 MeV and determine the $\beta_{2}$, the small value of which is indicative of the spherical closed-shell character of $^{24}$O.~A comparison of the $E_{\rm x}(2_1^+)$ and $\beta_{2}$ for the chain of oxygen isotopes shows strong evidence for a large shell gap at $N$ = 16.

\begin{figure}[t]
\begin{center}
\resizebox{1\columnwidth}{!}{%
\includegraphics{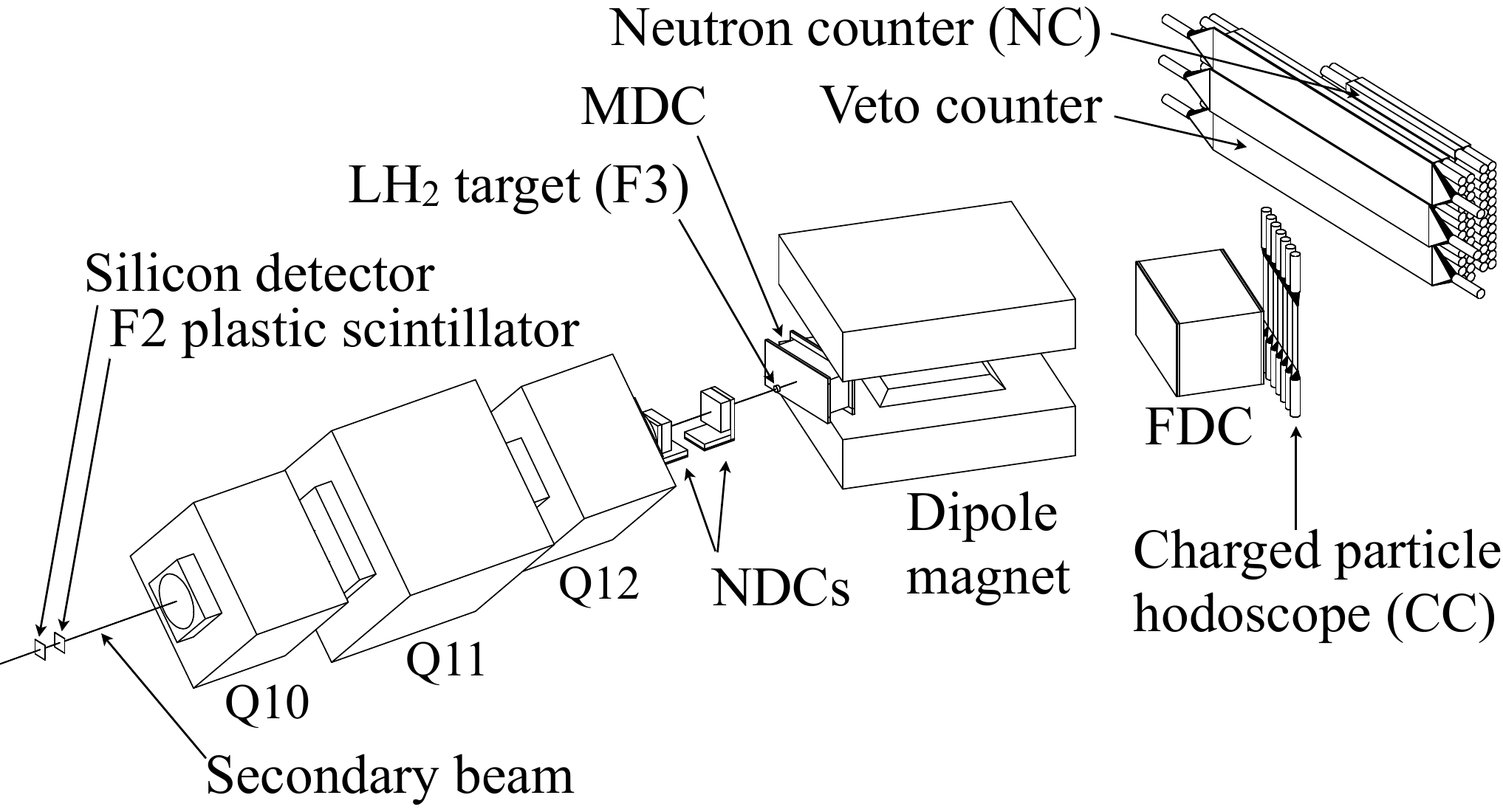}%
}
\caption{A schematic view of the experimental setup.
\label{fig1}}
% \vspace{-10pt}
 \end{center}
%\end{figure}
%%%%%%%%%%%%%%%%%%%%%%%%%%%%%%%%%%%%%%%%%%%%%%%%%%%%%%
%%%%%%%%%%%%%%%%%%%%%%%%FIG. 2%%%%%%%%%%%%%%%%%%%%%%%%
%\begin{figure}[b]
\begin{center}
\resizebox{1\columnwidth}{!}{%
\includegraphics{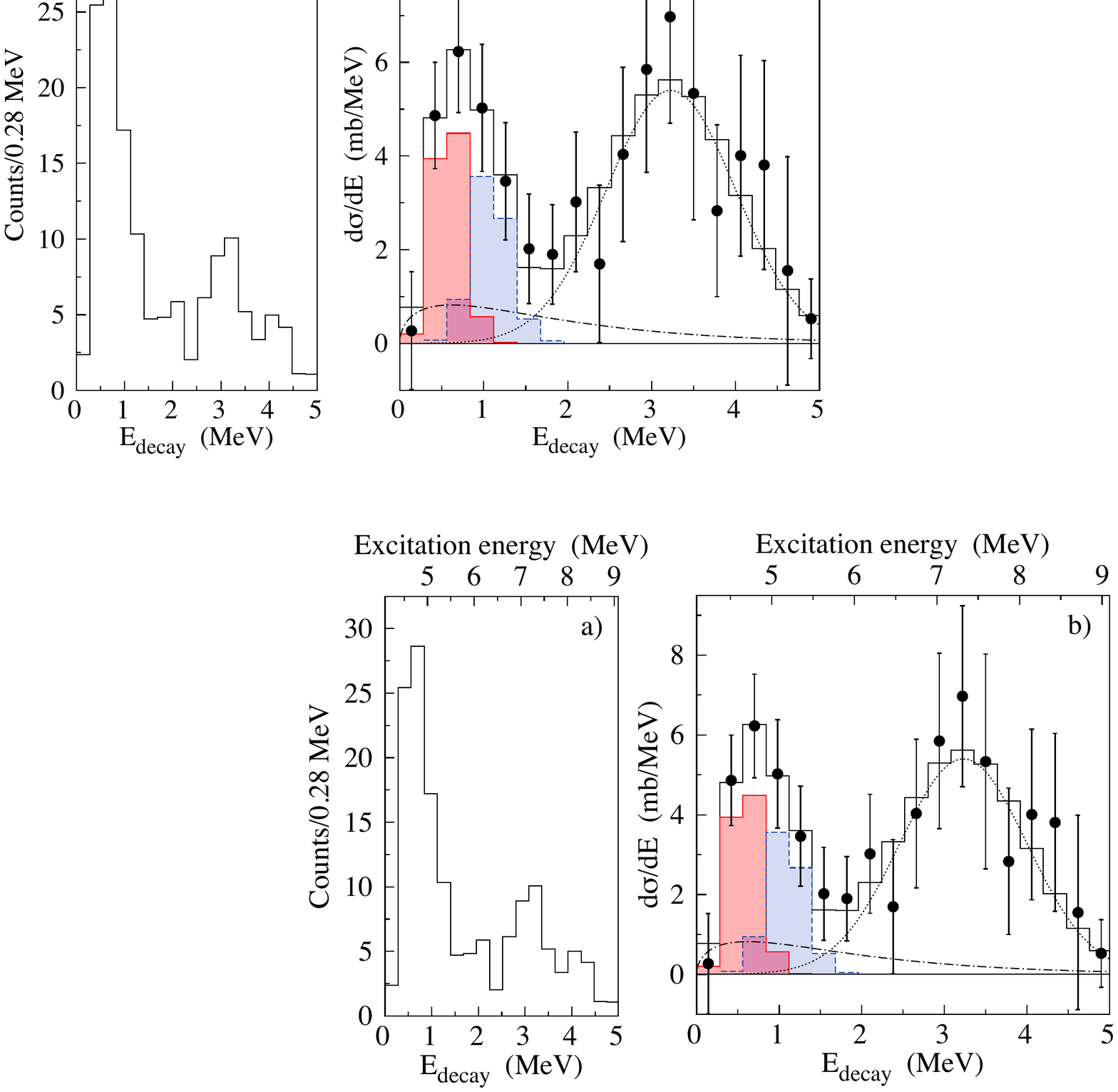}%
}
\caption{(Color online) (a) The coincidence yield of $^{23}$O and a neutron in decay energy and (b) the cross section, d$\sigma$/d$E_{\rm decay}$.~The error bars are statistical only.~The red-solid and blue-dashed histograms are the results of fits for resonances at $E_{\rm decay }$ = 0.56 MeV and 1.06 MeV, respectively.~The dotted Gaussian function at $E_{\rm decay }$ $\sim$ 3.2 MeV represents higher-lying state(s).~The dot-dashed line corresponds to the non-resonant continuum~\cite{DEAK}.
\label{fig2}}
 \vspace{-20pt}
 \end{center}
\end{figure}

The experiment was performed at the RIPS facility~\cite{Kubo} at RIKEN.~A schematic view of the downstream section of RIPS and the experimental setup is shown in Fig.~\ref{fig1}.~The $^{24}$O secondary beam was produced using a 1.5 mm-thick Be production target and a 95 MeV/nucleon $^{40}$Ar primary beam with a typical intensity of 40 pnA.~The intensity of the $^{24}$O secondary beam was $\sim$4 ions/sec with a momentum spread of $\Delta$$p$/$p$ = $\pm$ 3\%.~The liquid-hydrogen (LH$_{2}$) target~\cite{LH2} was installed at the achromatic focus F3 of RIPS.~The effective target thickness and the mid-target energy were 159 $\pm$ 3 mg/cm$^2$ and 62 MeV/nucleon, respectively.~The secondary beam was tracked particle-by-particle on to the target using two drift chambers (NDCs) located just upstream of the target.~The $B$$\rho$-TOF-$\Delta$$E$ method was employed to identify the charged fragments following reactions of the $^{24}$O beam with the LH$_2$ target.~The magnetic rigidity ($B\rho$) was determined from the position and angle measurements derived from the MDC and FDC drift chambers (Fig.~\ref{fig1}).~The TOF (time-of-flight) of the fragments was measured between the target and the plastic scintillator charged particle hodoscope (CC).~The energy loss ($\Delta E$) was measured using the CC.

The neutrons were detected using the plastic scintillator neutron counter (NC) array placed some 4.7 m downstream of the target together with the charged particle veto counter (Fig.~\ref{fig1}).~The NC array consisted of four layers with a total thickness of 24.4 cm.~The momentum vectors of the neutrons were determined from the TOF between the target and the NC along with the hit positions at the target and the neutron counter.~A neutron detection efficiency of 25.0 $\pm$ 0.8$\%$ at 64 MeV was measured for a 2 MeVee threshold in a separate $^7$Li($p,n$) run.

The decay energy spectrum of $^{24}$O$^*$ was reconstructed from the measured four momenta of $^{23}$O and the emitted neutron.~The decay energy, $E_{\rm decay}$, is expressed as:
\begin{equation}
E_{\rm decay}=\sqrt{ (E_f + E_n)^2-|\boldsymbol{p}_f +\boldsymbol{p}_n|^2}-(M_f+M_n)~, 
\label{eq:proportionality_1}
\end{equation}
where $E_f (E_n)$ and $\boldsymbol{p}_f (\boldsymbol{p}_n)$ are the total energy and the momentum of $^{23}$O (neutron) and $M_f $ and $M_n$ are the masses of $^{23}$O and the neutron, respectively.

%
%================== ANALYSIS & DISCUSSION ==============
%%%%%%%%%%%%%%%%%%%%%%%%% FIG 1 %%%%%%%%%%%%%%%%%%%
%%%%%%%%%%%%%%%%%%%%%%%%%%%%%%%%%%%%%%%%%%%%%%%%%%%%%%%
%

%------------------------------ decay energy spectrum raw ---------------------------
%Figure~\ref{fig2}(a) shows the coincidence yield for $^{23}$O and a neutron in decay energy, obtained after subtracting contaminations from nearby $^{22,24}$O fragments, which existed due to limited resolution in fragment mass.~Two peaks are clearly visible at $E_{\rm decay}$ $\sim$ 0.7 and $\sim$3.2 MeV.
Figure~\ref{fig2}(a) shows the coincidence yield for $^{23}$O and a neutron in decay energy.~Two peaks are clearly visible at $E_{\rm decay}$ $\sim$ 0.7 and $\sim$3.2 MeV.~Figure~\ref{fig2}(b) shows the decay energy spectrum in terms of cross section (d$\sigma$/d$E_{\rm decay}$) after correction for the detection efficiencies and acceptances.~The error bars are statistical only.~The geometrical acceptance was estimated using a Monte Carlo simulation taking into account the beam profile, the geometry of the setup, the experimental resolutions, and the multiple scattering of the charged particles.~The acceptance of the neutron detector drops rapidly as the decay energy increases, resulting in a significant suppression of the higher-lying strength.

%------------------------------ two resonances and high-lying state ---------------------------
The decay energy spectrum was fitted using two resonance distributions for the first peak at $E_{\rm decay}$ $\sim$ 0.7 MeV, a Gaussian distribution for the broad feature at $E_{\rm decay}$ $\sim$ 3.2 MeV, and a Maxwell distribution for the non-resonant continuum background~\cite{DEAK}.~The first peak around $\sim$0.7 MeV was interpreted as two closely spaced resonances because the width is greater than the estimated experimental energy resolution by a factor of $\sim$2.~The experimental energy resolution is energy dependent and estimated to be $\Delta E_{\rm decay}$ $\approx$ 0.5$\sqrt{E_{\rm decay}}$ (MeV) in FWHM.~The single-particle width for a $d$-wave neutron resonance is predicted to be less than $\sim$0.1 MeV~\cite{Bohr}.~Therefore, the experimental resolutions were adopted for the widths.~The solid histogram of Fig.~\ref{fig2}(b) represents the best fit result of the overall distribution of the total cross section, d$\sigma$/d$E_{\rm decay}$, which consists of two resonances of the first peak, a Gaussian distribution for the higher lying peak at $\sim$3.7 MeV and the non-resonant continuum background.~The red-solid and blue-dashed histograms are those for the first two resonance states at $E_{\rm decay }$ = 0.56 $\pm$ 0.05 MeV and 1.06 $\pm$ 0.10 MeV, which correspond to $E_{\rm x}$ = 4.65 $\pm$ 0.14 MeV and 5.15 $\pm$ 0.16 MeV, respectively, adopting the separation energy $S_n$ = 4.09 $\pm$ 0.13 MeV~\cite{Jurado}.~The dot dashed line is the distribution for the non-resonant continuum background obtained from the same fit.~The excitation energy of the first resonance is consistent with that of the recent study of Hoffman \textit{et al.}~(4.72 $\pm$ 0.11 MeV)~\cite{Hoffman09} and in accord with the $\nu1s_{1/2}-\nu0d_{3/2}$ shell gap (4.86 $\pm$ 0.13 MeV) derived from the location of the $^{25}$O$_{\rm g.s.}$ resonance~\cite{Hoffman08}.

\begin{figure}[t]
\begin{center}
\resizebox{1\columnwidth}{!}{%
\includegraphics[width=8.8cm]{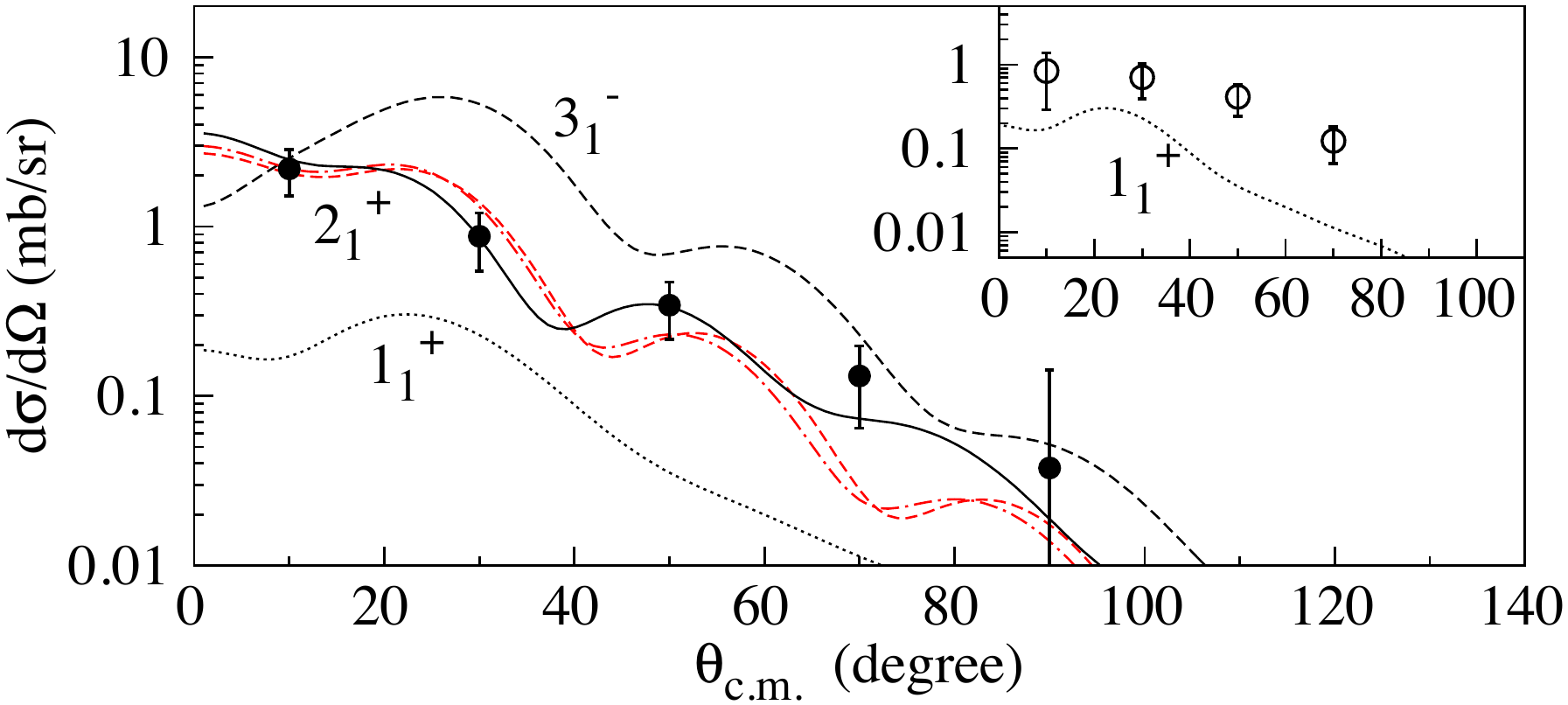}%
}
\caption{(Color online) Angular distributions for the resonances at $E_{\rm decay }$ = 0.56 (filled circles) and 1.06 MeV (open circles in the inset).~The error bars are statistical only.~The curves represent the results of microscopic (black curves) and phenomenological (red curves) DWBA calculations (see text).
\label{fig3}}
\end{center}
%\end{figure}
%\begin{figure}[t]
\begin{center}
\resizebox{1\columnwidth}{!}{%
\includegraphics[width=8.3cm]{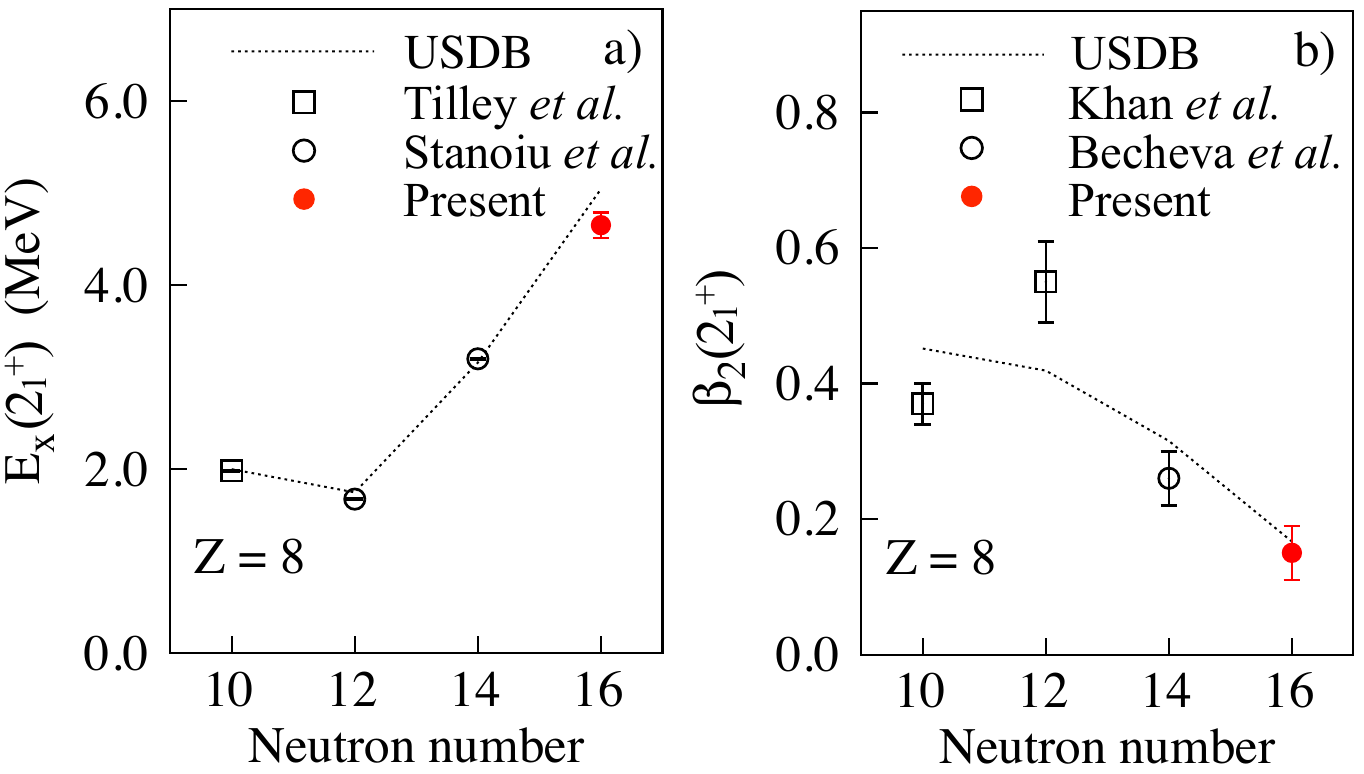}%
}
\caption{(Color online) The neutron number dependences of (a) $E_{\rm x}$(2$_1^+$) and (b) $\beta_{2}$(2$_1^+$).~The present results are shown by the filled circles.~The $E_{\rm x}$(2$_1^+$) for $N$= 10$-$14 were taken from Refs.~\cite{Stanoiu04,Tilley95}.~The $\beta_{2}$ for $N$= 10$-$14 were taken from Refs.~\cite{Khan00,Becheva06}.~The dotted lines represent the USDB shell model predictions~\cite{USDab}.
\label{fig4}}
 \vspace{-20pt}
 \end{center}
\end{figure} 
%%%%%%%%%%%%%%%%%%%%

The dotted line in Fig.~\ref{fig2}(b) shows the fit to the high-lying peak with $E_{\rm decay}$ $\approx$ 3.2 MeV corresponding to an excitation energy of 7.3 MeV.~Recently a high-lying state has been reported at $E_{\rm x}$ $\approx$ 7.5 MeV, which was produced via proton knock-out from $^{26}$F and de-excited to $^{22}$O$_{\rm g.s.}$ via two neutron cascade, $^{22}$O$_{\rm g.s.}$+2$n$~\cite{Hoff11}.~This is a different decay channel from the present one, $^{23}$O$_{\rm g.s.}$+$n$.~Since the decay channel is different, the strength observed here is additional to that previously reported, thereby implying considerable strength in the excitation energy region around 7.3 MeV.~The nature of this high-lying strength will be discussed later.

%--------------------------------- total cross section ------------------------------------
The total inelastic cross sections to the low-lying resonances at $E_{\rm decay }$ = 0.56 and 1.06 MeV were determined to be $\sigma$ = 2.6 $\pm$ 1.1 mb and 2.2 $\pm$ 1.2 mb, respectively, after subtracting the non-resonant continuum.~The quoted errors mainly come from the uncertainties in the fitting ($\sim$39\% and $\sim$47\%) and the choice of the functional form describing the non-resonant continuum ($\sim$19\% and $\sim$27\%).

%--------------------------------- single particle configuration ------------------------------------
The single-particle configurations for the first and second excited states of $^{24}$O are expected to be mainly $\nu(1s_{1/2})^{-1}\nu(0d_{3/2})^1$ $J^{\pi}$ = 1$^+$ or 2$^+$ in the shell model picture.~Calculations using the universal $sd$-shell model interactions, USD~\cite{USD} and USDA/B~\cite{USDab}, as well as that including the continuum states~\cite{Alexander05} all predict that the 2$^+$ state is lower lying than the 1$^+$ state by approximately 0.5$-$1.0 MeV.

%--------------------------------- Angular distribution ------------------------------------
%The angular distributions of the cross sections were obtained by fitting the decay energy spectra for each angular bin, assuming that the non-resonant continuum is uniform in the angular distribution.~The angular distributions (for angle bins of 20$^\circ$) are displayed in Fig.~\ref{fig3}.

The angular distribution, d$\sigma$/d$\Omega$, for the first resonance state of Fig.~\ref{fig3} was obtained by fitting the decay energy spectrum of each angular bin (of 20$^\circ$) in the similar way to the fitting method explained for Fig.~\ref{fig2}(b) but adopting the uniform angular distribution for the non-resonant continuum background.~With this the decay energy spectrum at each angle could be fitted with only the resonance distributions after the subtraction of the non-resonant background whose total cross section was fixed in the explained procedure for Fig.~\ref{fig2}(b).~The error bars are statistical only.~The experimentally determined angular distributions have been compared with microscopic DWBA calculations performed using the {\sc dw81} code~\cite{DW81}.~The global optical potential KD02~\cite{KD} and the M3Y~\cite{M3Y} were employed for the distorted wave function and for the effective nucleon-nucleon interaction, respectively.~The one-body transition densities were calculated using the shell-model code {\sc nushell}~\cite{Nushell}.~The calculations used the USDB interaction for the 2$_1^+$ and the 1$_1^+$ states, and the WBT~\cite{WBT} interaction for the 3$_1^-$ state.~The size parameter $b$ = 2.03 $\pm$ 0.08 fm was chosen to reproduce the rms radius of $^{24}$O~\cite{OzawaNPA01} within the harmonic oscillator potential.~The results of the calculation for the transitions 0$^+_{\rm g.s.}$$\rightarrow$$2_1^+$ (black-solid line), 0$^+_{\rm g.s.}$$\rightarrow$$1_1^+$ (black-dotted line), and 0$^+_{\rm g.s.}$$\rightarrow$$3_1^-$ (black-dashed line) are shown in Fig.~\ref{fig3}.~It may be noted that there are no adjustable normalization parameters in the calculation.~The result for the 0$^+_{\rm g.s.}$$\rightarrow$$2_1^+$ $\{\nu(1s_{1/2})^{-1}\nu(0d_{3/2})^1\}$ transition reproduces the angular distribution of the resonance at $E_{\rm decay}$ = 0.56 MeV very well, strongly supporting a spin-parity assignment $J^{\pi}$ = 2$^+$.

%--------------------------------- collective DWBA ------------------------------------
%Next the $\beta_{2}$ is derived by comparing the total cross section with that of a phenomenological collective DWBA calculation.~The DWBA cross section for the $2_1^+$ state was normalized to the measured total cross section with $\beta_{2}$ = 0.15 $\pm$ 0.04, which is the average of those for the two different optical potentials.~The red-dashed and red-dot-dashed lines in Fig.~\ref{fig3} represent the results of the phenomenological DWBA calculations using the {\sc ecis97} code~\cite{ECIS97} for two different global optical potentials KD02 and CH89~\cite{CH89}, respectively.~The error in the $\beta_{2}$ reflects the uncertainty in the total cross section and the choice of optical parameters.

The quadrupole transition parameter $\beta_{2}$ is a measure of the deformation~\cite{Becheva06}.~We have derived the $\beta_{2}$ value of the 2$^+_1$ state of $^{24}$O by normalizing the phenomenological collective DWBA calculation to the measured total cross section.~The red-dashed and red-dot-dashed lines in Fig.~\ref{fig3} represent the normalized angular distributions of the collective DWBA calculations using the {\sc ecis97} code~\cite{ECIS97} for the two different optical potentials KD02 and CH89~\cite{CH89}, respectively, from which we deduce $\beta_{2}$ = 0.15 $\pm$ 0.04.~The error reflects the uncertainty in the total cross section and the choice of optical parameters.

%--------------------------------- systematics ------------------------------------
%Turning now to the systematics Fig.~\ref{fig4} shows the dependence in neutron number of $E_{\rm x}(2_1^+)$ and $\beta_{2}$($2_1^+$) for the even-even oxygen isotopes.~It is clear in Fig.~\ref{fig4}(a) that the excitation energies from the USDB shell model calculations reproduce well the experimental $E_{\rm x}(2_1^+)$.~Moreover, the $E_{\rm x}(2_1^+)$ increase considerably in moving from $N$ = 12 to 16.~Based on the $\beta_{2}$ = 0.26 $\pm$ 0.04 derived from proton inelastic scattering, Becheva $et~al$.~demonstrated that the $\nu0d_{5/2}$ sub-shell is closed in $^{22}$O ($N$ = 14).~In the case of $^{24}$O ($N$ = 16) the 2$^+_1$ state lies some 1.5 MeV higher than in $^{22}$O, while the $\beta_{2}$ (0.15 $\pm$ 0.04) is smaller by around 0.1.~As such, the gap at $N$ = 16 must be considerably larger than at $N$ = 14.~This trend is in accord with the predictions of Otsuka $et~al$.~\cite{Otsuka01,Otsuka05,Otsuka10} whereby the $\nu1s_{1/2}$ sub-shell is closed in $^{24}$O.

Figure~\ref{fig4} shows the neutron number dependences of the measured and calculated (dotted lines) $E_{\rm x}(2_1^+)$ and $\beta_{2}$($2_1^+$) for the even-even oxygen isotopes.~The $E_{\rm x}(2_1^+)$ values increase considerably in moving from $N$ = 12 to 16.~This trend and the high excitation energy at $N$ = 16 are well reproduced by the USDB shell model calculations.~In the case of $^{24}$O ($N$ = 16), the 2$^+_1$ state lies $\sim$1.5 MeV higher than in $^{22}$O, reflecting the large gap between the $\nu1s_{1/2}$ and $\nu0d_{3/2}$ orbitals, which is a maximum at $N$ = 16.

The dotted line in Fig.~\ref{fig4}(b) represents the USDB shell model predictions using effective charges of $e_p$ = 1.36 and $e_n$ = 0.45~\cite{RichterEffC} and following Bernstein's prescription~\cite{Bernstein}.~The calculations reproduce reasonably well the experimental $\beta_{2}$ values, in particular those of $^{22}$O and $^{24}$O.~Based on the small $\beta_{2}$ value (0.26 $\pm$ 0.04) derived from the proton inelastic scattering of $^{22}$O, Becheva $et~al$.~concluded that the $\nu0d_{5/2}$ sub-shell is closed at $N$ = 14.~For $^{24}$O, the $\beta_{2}$ value (0.15 $\pm$ 0.04) is even smaller than that of $^{22}$O, thereby suggesting that it is the least deformed of the oxygen isotopes.~From the $E_{\rm x}(2_1^+)$ and $\beta_{2}$ values of $^{22}$O and $^{24}$O, it is clear that the gap at $N$ = 16 must be considerably larger than at $N$ = 14.~This is in accord with the predictions of Otsuka $et~al$.~\cite{Otsuka01,Otsuka05,Otsuka10} whereby the $\nu1s_{1/2}$ sub-shell is closed in $^{24}$O.

%------------------------------ low-lying negative parity state --------------------------------
While the cross section for the first excited state is well reproduced by the microscopic DWBA calculations for the 0$^+_{\rm g.s.}$$\rightarrow$2$_1^+$ $\{\nu(1s_{1/2})^{-1}\nu(0d_{3/2})^1\}$ transition, that of the second resonance, 2.2 $\pm$ 1.2 mb, is much larger than that calculated for the 0$^+_{\rm g.s.}$$\rightarrow$1$_1^+$ $\{\nu(1s_{1/2})^{-1}\nu(0d_{3/2})^1\}$ transition (Fig.~\ref{fig3}).~While the uncertainty in the measured cross section (56$\%$) is large, one conjecture is that some of the missing strength might be attributed to negative-parity states $\nu(1s_{1/2})^{-1}\nu(fp)^1$ that come down in energy to lie near the $2_1^+$ state owing to the quenching of the gap between the $\nu0d_{3/2}$ orbital and the $\nu fp$ shell.~Indeed, low-lying intruder 3/2$_1^-$ and 7/2$_1^-$states in $^{27}$Ne~\cite{SMBrown} indicate a narrowing of the gap between the $\nu0d_{3/2}$ orbital and the $\nu fp$ shell.

%------------------------------ high-lying negative parity state --------------------------------
As mentioned earlier, the high-lying strength ($E_{\rm x}$ $\sim$ 7.3 MeV) observed here represents an additional contribution to that observed previously in the two-neutron emission channel~\cite{Hoff11}.~High-lying states produced via the promotion of a $0d_{5/2}$ neutron would decay most likely to the 5/2$^+$ first excited state of $^{23}$O$-$a $\nu(0d_{5/2})^{-1}$ hole state$-$which is unbound~\cite{Schiller07,SatouFBS} and, in turn, decays to $^{22}$O$_{\rm g.s.}$ via neutron emission, a process not observable in the present experiment.~This implies that the states observed here around $\sim$7.3 MeV would have mostly $\nu(1s_{1/2})^{-1}\nu(fp)^1$ negative-parity configurations.

%======================= SUMMARY =======================
In summary, we have investigated the unbound excited states of neutron-rich $^{24}$O via proton inelastic scattering in inverse kinematics.~The excitation energy of the first excited state was determined to be 4.65 $\pm$ 0.14 MeV and the spin-parity was assigned $J^{\pi}$ = 2$^+$.~In addition, a relatively small $\beta_{2}$ parameter was determined (0.15 $\pm$ 0.04), indicative of the spherical character of $^{24}$O and the large shell gap at $N$ = 16, confirming theoretical predictions~\cite{Otsuka01,Otsuka05,Otsuka10,BrownRichter}.~Finally, the strong higher-lying strength, identified as negative-parity $\nu(1s_{1/2})^{-1}\nu(fp)^1$ excitations, is suggestive of the quenching of the gap between the neutron $sd$ and $fp$ shells, as observed in $^{27}$Ne~\cite{SMBrown}.

%%%%%%%%%%%%%%%%%% acknowledgments %%%%%%%%%%%%%%%%%%

\begin{acknowledgments}
We would like to thank the accelerator operations staff of RIKEN for providing the $^{40}$Ar beam.~This work is supported by the Grant-in-Aid for Scientific Research (No.~19740133) from MEXT Japan, the WCU program and Grant 2010-0024521 of the NRF Korea.
\end{acknowledgments}


\end{document}